# eQETIC: A Maturity Model for Online Education

*Rogério Rossi & Pollyana Notargiacomo Mustaro*
*University of São Paulo and Mackenzie Presbyterian University, São Paulo, Brazil*

rossirogerio@hotmail.com   pollyana.mustaro@mackenzie.br

## Abstract

Digital solutions have substantially contributed to the growth and dissemination of education. The distance education modality has been presented as an opportunity for worldwide students in many types of courses. However, projects of digital educational platforms require different expertise including knowledge areas such as pedagogy, psychology, computing, and digital technologies associated with education that allow the correct development and application of these solutions. To support the evolution of such solutions with satisfactory quality indicators, this research presents a model focused on quality of online educational solutions grounded in an approach aimed to continuous process improvement. The model considers of three maturity levels and six common entities that address the specific practices for planning and developing digital educational solutions, targeting quality standards that satisfy their users, such as students, teachers, tutors, and other people involved in development and use of these kinds of educational solutions.

## Introduction

Digital technologies that drive social activities enable evolution in several areas, such as medicine, entertainment, industrial automation, and education. Specifically for education, this digital capability has enabled the availability of many products and services, i.e., digital solutions, involving an agile growth of educational ability, especially when it comes to online education.

It is possible to verify an expansive development for online education that considers the availability of automated instruction in courses offered in the distance modality and also according to the diversity of learning objects, developed using Information and Communication Technologies (ICT).

According to surveys planned in an info graphic about digital education presented by Knewton (2012), 95% of teachers felt that the engagement of students increased when they combined some type of digital technology with instruction transmission. Some projections presented herein estimate that, in 2020, at least 98% of the courses will be constituted of hybrid solutions, i.e., they will consider traditional learning approaches associated with online digital components.

According to the Brazilian federal government reports presented by MEC/INEP (2009), a significant increase has been observed in the registration of online courses in distance mode in Brazil. The report shows that 107 courses in online mode were registered in 2004 and that the number of registered courses jumped to 647 in 2008, which represents a high growth rate of courses for this modality, 605% in five years.

This growth has required some mechanisms and criteria for evaluating educational products and services based on digital technologies from control associations and from governments, specifically by departments concerned with educational control, at all levels and in all continents. These evaluation mechanisms should meet the requirements of certification or accreditation for such products; however, they should also seek to assess the quality of these products, which implies careful studies to use a method to evaluate the quality of digital educational solutions.

For Rekkedal (2006), contemporary society requires better quality from the many products and services offered, besides demanding this for online education, offered in its varied forms. To meet this kind of demand in the educational sector, governments, associations, universities, researchers, and other related groups are increasingly engaged in formalizing mechanisms and criteria to assess the quality of this kind of solutions as shown by some examples available in IHEP (2000), ISO (2009), and MEC/SEED (2007) among others.

Thus, for technically observing the quality of such solutions, specific mechanisms should be considered to encourage their construction and subsequent measurement, generating objective results that can be observed uniformly. For constructing and evaluating digital educational products, some specific frameworks can be considered, such as models, standards, processes, and many types of tools, as well as insightful measurement mechanisms to evaluate the products, specifically the entire digital solutions and their related development processes.

Therefore, this paper aims to present a quality model for digital educational solutions, called eQETIC (Quality Model for Educational Products based on Information and Communication Technology). The model considers an approach of continuous processes improvement and the subsequent maturity for quality assurance. This approach seeks to favor the desired quality criteria and the mechanisms for quality construction.

To meet such objectives, this article is organized as follows: section two presents a literature review focused on digital educational solutions and specific concepts and frameworks for digital education; section three considers works related to this research that also have models directed to the quality of digital educational solutions; section four discusses the criteria used for defining the eQETIC Model; section five analyzes the structural features of the model and its application; lastly, section six presents the conclusion and proposals for further works.

# Quality in Online Education

## *Online Educational Products*

The digital technology capability has provided significant changes in many sectors of modern society. In education, these changes specifically promote the wide dissemination of digital educational solutions that have different objectives and ways of implementation, for different devices and different communication channels.

Various designations and terminology concerning these products are presented by Guri-Rosenblit (2005), Piva Júnior, Pupo, Gamez & Oliveira (2011), and Perkins (2008). According to Guri-Rosenblit (2005), the following terminology is identified to refer to such products: 'distributed learning', 'online learning', 'web-based learning', 'virtual classrooms', 'online education' and 'I-Campus'.

Guri-Rosenblit (2005) highlights that two of the most widely used terms have conceptual differences, i.e., distance education and e-learning. Perkins (2008) confirms this difference, and many other authors make this distinction. They consider that the terms distance education and e-learning overlap each other in some cases, but do not have identical meanings. To Moore and Kearsley (2011), distance education corresponds to the planned learning that normally occurs away from the teaching place, requiring specific techniques for course creation through various technologies and special organizational and administrative provisions. For Barker (2007), e-learning represents an environment where learning occurs using both computers and the Internet.

Considered a recent phenomenon, e-learning is related to the use of electronic media for different learning proposals that can happen in the classroom environment and even replace the face-to-face virtual meetings, as mentioned by Guri-Rosenblit (2005). Perkins (2008) considers that e-learning can be roughly understood as digital technologies able to reduce costs and to improve learning.

## *Quality Concepts for Online Education*

Quality takes different concepts that can be considered to be a subjective characteristic, inherent to a product or service, but technically it should consider objective measures. Although a single, comprehensive, definition for the term is not found, there is a common sense to many authors, such as Crosby (1979) and Humphrey (1989), who consider quality as the compliance with the requirements.

For constructing online educational products, quality should be observed from different perspectives, considering criteria that address the psycho-pedagogical, social, technological issues, etc.

Garvin (1992) proposed what is called the four eras of quality, and these can be observed differently for each product category. The four quality eras correspond to inspection, quality control, quality assurance, and strategy to quality. The author suggests that when a product category is in the era of quality assurance, it means that the quality has to be built along the product development according to its defined process.

For the software industry, the approaches of quality control and quality assurance are considered and addressed by authors such as Pressman (2011) and Sommerville (2003). For software, Godbole (2005) is emphatic in stating that quality control and quality assurance are complementary approaches, yet quality assurance is a preventive approach, while quality control is a corrective approach.

In the context of online education, Barker (2007) considers that aspects related to quality are relevant for two reasons: first, because they are able to support buyers in their decisions to purchase products; and second, because they favor those who develop and offer these kinds of products.

For Pawlowski (2007), quality is appropriately meeting the stakeholders' goals and needs. In the context of e-learning, quality is related to all processes, products, and services for learning that are mediated by the use of information and communication technologies. This scenario can be tracked by online education frameworks.

## *Specific Frameworks for Online Education*

A diversity of frameworks has been proposed to manage the design, development, and maintenance of online educational products, sometimes with objectives concerning the certification of the products in a quality control approach.

Rekkedal (2006), Pawlowski (2007) and Shelton (2011) present a set of frameworks developed by researchers, associations, and governments that discuss the quality of online educational prod-eQETIC: A Maturity Model for Online Educationucts. Some examples are highlighted below, including quality frameworks designed by governments and specific quality organizations:

- IHEP's Quality on the line: Benchmarks for Success in Internet-Based Distance Education;
- QAA – Quality Assurance Agency for Higher Education: Guidelines on the Quality Assurance of Distance Learning;
- Open eQuality Learning Standards;
- Sloan consortium's five pillars of quality;
- ISO International Organization for Standardization – ISO/IEC 19796-1 Standard on Quality for e-learning; and
- MEC/SEED Quality Benchmarks for Distance Education.

These examples refer to frameworks that have been defined with different objectives and methods of use, which can be applied to various purposes.

However, it is also possible to verify a set of standards in MarylandOnline (2011) that can be used to measure online educational products, called QMRubric. This framework considers eight groups of standards that collaborate in the evaluation of online courses. This program was sponsored by a U.S. fund for developing a proposal to promote evaluation of online courses and quality control and to anticipate future issues of accreditation for online courses.

# Related Works

The software industry has developed models that favor the quality of the software product through continuous process improvement associated with its development. Maturity models, as they are known, generally favor the planning, development, acquisition, and maintenance of the software product and its components, thus being increasingly used by the industry in a larger scale.

Chrissis, Konrad, and Shrum (2004) present CMMI (Capability Maturity Model Integration), an example of a maturity model from the software industry. Nunes, Albernaz, and Nobre (2009) suggest the process improvement approach used by the software industry could also be replicated to define a process for digital educational solutions, such as distance education and e-learning, stimulating the development and quality of these products.

It is possible to verify related works that discuss the models which favor the quality of digital educational solutions focusing on a continuous process improvement approach, as can be seen in the eQETIC Model, which is the subject of this article. As an example, Marshall and Mitchell (2002) allow observing a proposed maturity model for e-learning called 'e-Learning Maturity Model' that considers five maturity levels: 1) Initial; 2) Planned; 3) Defined; 4) Managed; and 5) Optimization. The 'e-learning maturity model' proposed by the authors considers the same structure of maturity levels verified in the extinct model called SW-CMM (Software Capability Maturity Model).

Another model proposed by Khan (2004), called 'e-Learning P3 Model', considers people, process, and product, defining the following stages in its structure: 1) Planning, 2) Designing, 3) Production, 4) Assessment, 5) Delivery and Maintenance, 6) Instruction, and 7) Marketing.

Marshall and Mitchell (2002) also present a study aimed to apply the principles defined by the SPICE (Software Process Improvement and Capability dEtermination) project to develop e-learning processes, and in this study, the authors consider specific criteria for the following processes, 1) Learning, 2) Development, 3) Coordination and Support, 4) Evaluation, and 5) Organi-zation, which are measured according to the following parameters, 'not adequate, 'partially adequate', 'largely adequate' and 'totally adequate'.

The EduQNet was presented by Rapchan, Cury, Menezes, and Falbo (2002) to meet the requirements of quality for distance learning courses mediated by Internet. The authors considered the application of standard ISO / IEC 12.207 (Information Technology - Software Life Cycle Process) applied to digital educational solutions. The model is organized into fourteen main activities, and each of these indicates a set of sub-activities.

These works show that the continuous process improvement approaches are being applied to the construction of models for digital educational solutions. Considering the proposals presented by Marshall and Mitchell (2002) and Khan (2004), they are verified to be restricted to the e-learning product and the proposal identified in Rapchan et al. (2002) is restricted to distance educational courses mediated by the Internet.

Similarly to these works, the continuous process improvement approach was applied to the design and structure of the eQETIC Model, as evidenced in more detail in the following sections.

## Fundamentals of the eQETIC Model

Practices regarding development and construction of digital educational solutions are varied, as are the models and standards that support this activity. This highlights the need of structured mechanisms able to hold actions in a local, regional, or even in a global view.

The influences exerted on these solutions in their planning, development, and maintenance phase are diverse and, according to Hadjerrouit (2007), the construction of an e-learning product should include educational, organizational, pedagogical, and technological dimensions. This diversity of dimensions must be properly integrated in order to generate the best results.

Relevant characteristics of planning and development of courses in the distance education modality should consider the practices defined by instructional design theories. Tools and techniques associated with these theories assist the construction of such digital educational solutions giving a sense of pedagogical engineering.

The issues concerning the learning process are supported by the precepts of pedagogical engineering, but, besides these, there is an equally significant precept to be considered: the cognitive process. It must be effectively considered in the learning process in any of the learning modalities, including the modalities based on digital technologies (Gagné, Briggs, & Wager, 1992; West, Farmer, & Wolff, 1991).

Thus, the exploratory research used to develop the eQETIC Model (Quality Model for Educational Products Based on Information and Communication Technology) considers these influences, i.e., the pedagogical engineering through which the practices of instructional design and cognitive processes directed to learning can be verified.

Studies and researches have also resorted to review and detailed analysis of frameworks that exploit the quality, certification, accreditation, and development of general digital educational solutions.

Associated with these studies, the maturity model approach derived from Software Engineering contributed to the proposed model, as verified in the eQETIC Model structure, in which the principles of continuous process improvement and quality were used considering the integration of different dimensions related to digital educational solutions as in Rossi and Mustaro (2012).

These studies have provided critical insight and generated knowledge and ability that allowed defining the conceptual eQETIC model. Its structure was defined as well as its mode of applica-tion, guided by research into the models that consider the principles of continuous process improvement as presented in Rossi (2013).

Another key factor in the design of the model was the definition of which digital learning products could be considered by the model. In this sense, the model considers the courses offered in the distance education modality as one of the products: e-learning as another product, and learning objects as another kind of digital product. In this sense, some rules were created to be applied specifically to one product, but they can also be applied to the other products.

Considering the principles of continuous process improvement, three improvement levels were defined for the model, which are able to determine its degree of adherence and the degree to which an organization complies with the implementation rules defined by the model. The three improvement levels considered by the eQETIC Model are 1) Sufficient, 2) Intermediate, and 3) Global.

The improvement levels represent the view of continuous process improvement as they establish that institutional processes should be implemented and periodically improved to achieve the desired results. The three levels defined by the eQETIC Model consider the six common entities designated for the model that are repeated at each level and which have special rules for each of them. Each common entity comprises several implementation rules, unique to each level, and which favor the achievement of each level.

The common entities were based on studies and investigations that comprised the theoretical structure of this study, using book chapters, theses, scientific articles, and several frameworks issued by associations, governments, and researchers, from which are highlighted the following: 1) Khan's eight dimensions of e-learning framework; 2) IHEP's Quality on the line; 3) ISO/IEC 19.796-1 Standard of Quality for e-learning; 4) SLOAN Consortium five pillars of quality; and 5) NADE - Norwegian Association for Distance Education. Thus, each of the common entities considered by the model were defined with their respective function, with its set of rules and supported by specific references, as shown below:

**Didactic-Pedagogical Common Entity (DPCE)** - although the didactic and pedagogical issues are complex, they are not treated in depth by the model. Considering the context of this common entity related to applying modeling and construction of online courses, the theories of instructional design were considered, as well as the implications of cognitive processes in learning. To define this common entity, four of the main theoretical references were considered: West et al. (1991), Gagné et al. (1992), Dick, Carey, and Carey (2005), and Briggs (1977).

**Technology Common Entity (TECE)** - aims to define practices regarding the technological capacity of educational products. From the technology plan, the organization should possess software, security items, data storage, media, as well as the hardware infrastructure and telecommunication. The rules of this common entity were based on references such as Barker (2002), IHEP (2000), ENQA (2005), and MEC/SEED (2007).

**Management Common Entity (MACE)** - determines the management and operation ability to develop and to maintain digital educational solutions. Addresses strategic capability, project management and considers the implementation of quality system indicators to provide means to measure quality results, being supported by Martínez et al. (2011), Moore & Kearsley (2011), and PMI (2008).

**Support Common Entity (SUCE)** - considers the rules that determine the support mechanisms for learners involving all types of infrastructure, be it over the telephone, online, or in person. It also considers rules that provide support to tutors who may require detailed information to submit content-best results, and are considered in references such as IHEP (2000) and Colomina, Rochera, and Naranjo (2011).

**Tutorial Common Entity (TUCE)** - addresses the practical issues regarding the actions of mentoring in distance mode courses. Considers rules that address formal training, either regarding content or technological aspects that provide better performance to tutors. The practices were based on Elissavet and Economides (2003), Litto and Formiga (2009), and Pera, Cervera, and Barado (2007).

**Evaluation Common Entity (EVCE)** - includes considerations on the diagnostic, summative and formative assessment, and self-evaluation of the learning process. The rules of this entity consider ways of storing and disseminating evaluation and feedback controls that should be associated with the learning process. Online assessment tools are also considered in this entity, constructed based on West et al. (1991), Gagné et al. (1992), Dick et al. (2005), and Coll and Engel, (2011).

The Implementation Rules that belong to each of the common entities were also defined from these investigations that meet the needs for each of the observed improvement levels. The rules and other components that make up the structure of the model will be presented in detail in the next section.

## Structure and Application of the eQETIC Model

The structure overview of the eQETIC (Quality Model for Educational Products Based on Information and Communication Technology) can be seen in Figure 1. This structure is based on a continuous process improvement approach and considers three improvement levels, each of one including the six common entities (CE) defined for the model. Each of the CE considers a set of implementation rules (IR) that are grouped according to the groups of implementation rules (GIR). The GIR and IR defined are unique and not repeated over the model implying that a given IR defined for a specific CE is unique and exclusive to this entity at this level.

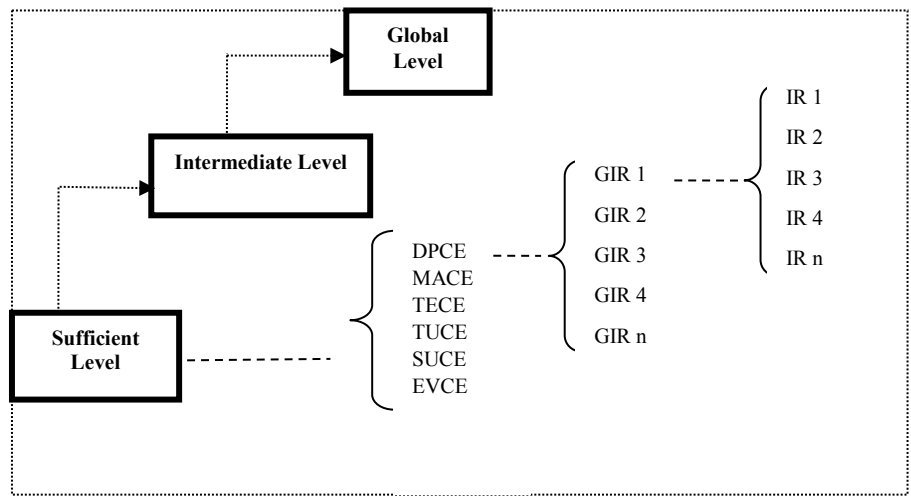

**Figure 1: eQETIC Model Structure (Rossi & Mustaro, 2012)**

The goal of each improvement level is highlighted below. It should be noted, however, that an organization that seeks to implement the eQETIC Model should follow the rules defined at each improvement level, implementing or adjusting their processes according to the IR of that level. Hence, after observing the rules of a given level, the subsequent levels should be met. The improvement levels considered by the model are:

- Sufficient level - allows sufficient functional condition and considers basic and fundamental implementation rules;
- Intermediate Level - allows improvement in processes according to the rules at this level, generating continuous product improvement in order to improve quality through process changes and adjustments;
- Global level - considers all the rules of the previous levels, as well as of this level, allowing global and full use of the model where the organization is considered to comply with the rules laid down therein.

The Common Entities (CE) aim to group the implementation rules that must be followed by the organization using the model. Six common entities are considered, as previously mentioned, that include rules capable of supporting the institutionalization of processes favoring the planning, development, and maintenance of digital educational solutions, as follows: Didactic-Pedagogical Common Entity (DPCE); Technology Common Entity (TECE); Management Common Entity (MACE); Support Common Entity (SUCE); Tutorial Common Entity (TUCE); Evaluation Common Entity (EVCE).

For managing its processes as prescribed by the eQETIC Model an organization must observe another element of the model that is not presented in Figure 1 and is totally relevant to the customization of eQETIC rules: the EPI (Educational Product Indicator).

Since the eQETIC Model provides rules for different ICT-based educational products, such as distance education, e-learning, and learning objects, the organization should consider a customization step of the model using the EPI. Each Implementation Rule has an associated EPI. The organization that aims to abide by the rules of the model must check which rules are relevant in accordance with the educational product. The EPI considers the following values: 'DE' for 'Distance Education' (emphasizing that, in this case, the model is suitable to the development and maintenance of courses); 'EL' which refers to 'e-learning ', and 'LO' that concerns Learning Objects.

Table 1 shows examples of Implementation Rules defined by the eQETIC Model, not restricted to a specific Improvement Level or Common Entity, just for presenting the rules as they are defined, grouped into their respective group, showing the associated EPI.

**Table 1: eQETIC Model – Examples of Implementation Rules (Rossi, 2013)**

| Improvement Level | Common Entity (1) | EPI (2) | Group of Implementation Rules (GIR) | Code/Description of Implementation Rules (IR) |
|---|---|---|---|---|
| 1 | DPCE | DE; EL | DPGIR 101 | DPIR 101.1 - Topics should be defined and documented |
| 1 | TECE | DE, EL | TEGIR 100 | TEIR 100.1 – A Technology Plan must be established and maintained. |
| 1 | MACE | DE | MAGIR 102 | MAIR 102.4 – A team of teachers and tutors should be defined. |
| 1 | EVCE | DE, EL | EVGIR 101 | EVIR 101.1 – The types of evaluation to be applied to the learner should be relevant to the nature of the content presented. |
| 2 | EVCE | ED, EL, LO | EVGIR 201 | EVIR 201.1 – Online mechanisms of learner self-assessment must be defined and implemented. |
| 3 | EVCE | DE, EL | EVGIR 301 | EVIR 301.1 – General evaluation mechanisms of online learning environment supported by Central Technological Systems must be established and maintained. |

Legend (1): 'DPCE – Didactic-Pedagogical Common Entity'; 'TECE – Technology Common Entity'; 'MACE – Management Common Entity'; 'EVCE – Evaluation Common Entity '.

Legend (2): 'DE – Distance Education'; 'EL – e-Learning'; 'LO – Learning Objects'.

As an example to illustrate the application of EPI, rule DPIR 101.1 (first line of Table 1) concerns only 'Distance Education' and 'e-learning' products; this rule is hence not considered for the 'Learning Object' product. The model presents rules that can sometimes be relevant to a unique product, as rule MAIR 102.4; and sometimes other rules pertain to the three products, such as rule EVIR 201.1 (fifth line of Table 1).

The Group of Implementation Rules (GIR) is another model component, as shown in Figure 1. The GIR aims to group a set of implementation rules that are totally related, allowing better understanding of the purpose of a particular group. Considering as an example rule TEIR 100.1 shown in Table 1, this rule belongs to the group TEGIR 100, which, according to eQETIC Model, refers to the process that determines that a Technology Plan should be defined by the organization. Group TEGIR 100, for example, considers two implementation rules in its totality.

Table 1 allows observing some examples of Implementation Rules (IR) defined by the model that define which should be implemented, i.e., what the institutionalized process should consider to be adherent to the eQETIC Model. As mentioned, all the Implementation Rules are unique and are associated with a Group of Implementation Rules (GIR), a Common Entity at a given Improvement Level of the eQETIC Model. According to Rossi (2013), the implementation rules represent the essence of the model being able to declare what the organization should consider for its institutional processes. Although only some Implementation Rules are highlighted in Table 1, the model has a total of 89 IR distributed among three improvement levels, as detailed in Table 2.

Table 2: eQETIC Model – Quantity of Implementation Rules

| Improvement Level | Common Entity | # GIR | # IR |
|---|---|---|---|
| Sufficient Level (1) | DPCE | 9 | 13 |
| | TECE | 5 | 9 |
| | MACE | 6 | 17 |
| | SUCE | 3 | 6 |
| | TUCE | 4 | 7 |
| | EVCE | 3 | 5 |
| Intermediate Level (2) | DPCE | 3 | 4 |
| | TECE | 1 | 1 |
| | MACE | 2 | 7 |
| | SUCE | 1 | 1 |
| | TUCE | 1 | 1 |
| | EVCE | 3 | 4 |
| Global Level (3) | DPCE | 2 | 2 |
| | TECE | 1 | 2 |
| | MACE | 2 | 4 |
| | SUCE | 1 | 1 |
| | TUCE | 1 | 1 |
| | EVCE | 2 | 4 |

With the structural view of the model, it is possible to emphasize that it is able to outline the process implementation in an organization that uses one of the online educational products consid-ered by the model. However, the organization should consider the rules that are established by the model in their processes, favoring their continuous improvement in order to build and to improve the quality of products and services.

The organization should pay particular attention to the measurement processes creating appropriate indicators to assess their results. This is relevant to a quality system and is also provided by eQETIC Model through one of the implementation rules of the Management Common Entity, which defines that the organization should establish a measurement process. With this process, the organization should provide the results measured, comparing them and presenting them for disseminating the actions related to quality.

## Conclusion and Further Works

The goal of developing a model capable of supporting steps that guide the planning, development, and maintenance of digital educational solutions was achieved by presenting the structure of eQETIC model. This model follows a continuous process improvement approach, whereas the implementation of processes in a developer organization of these types of solutions favors the development lifecycle and the quality of these solutions.

Presenting three improvement levels as maturity levels, the model allows the organization to implement the processes belonging to each level at a given time, and these levels and processes are organized in six common entities.

The model was built upon fundamental concepts of digital technologies related to education. It presents relevant principles regarding planning and development of digital educational solutions, as well as practices to operationalize organizations that seek to develop these types of solutions with better quality.

Further works associated with this may contribute to defining a model to assess and to certify the institutions using eQETIC Model. The proposal is implementing organizational processes for developing and maintaining digital educational solutions. It is feasible to define a specific framework to establish evaluation mechanisms considering activities involved in the certification process as rules and instruments to be used in the evaluation of organizations that use and apply the eQETIC Model.

## References

Barker, K. C. (2002). Canadian recommended e-learning guidelines. *FuturEd for Canadian Association for Community Education and Office of Learning Technologies,* 1-11. Retrieved October 19, 2014, from http://www.futured.com/pdf/CanREGs%20Eng.pdf

Barker, K. C. (2007). E-learning quality standards for consumer protection and consumer confidence: A Canadian case study in e-learning quality assurance. *Educational Technology & Society*, *10*, 109-119.

Briggs, L. J. (1977). *Instructional design*. New Jersey, USA: Educational Technology Publications.

Chrissis, M. B., Konrad, M., & Shrum, S. (2004). *CMMI: Guidelines for process integration and product improvement*. Boston, MA: Pearson Education.

Coll, C., & Engel, A. (2011). La calidad de lós materiales educativos multimedia: Dimensiones, indicadores y pautas para su análisis y valoración [The quality of multimedia educational materials: Dimensions, measures and guidelines for analysis and assessment]. In E. Barberá, T. Mauri, & J. Onrubia (Eds.), *Cómo valorar la calidad de la enseñanza basada en las TIC: Pautas e instrumentos de análisis* [*How to assess the quality of education based on ICT: guidelines and analysis tools*]. Barcelona: Editorial GRAÓ.

Colomina, R., Rochera, M. J., & Naranjo, M. (2011). La perspectiva de los usuários sobre la calidad de los materiales educativos multimedia y los procesos formativos em línea: Usos, utilidad y valoración [The perspective of the users concerning the quality of multimedia educational materials and training processes online: Applications, utility and valuation]. In: E. Barberá, T. Mauri, & J. Onrubia (Eds.), *Cómo valorar la calidad de la enseñanza basada en las TIC: Pautas e instrumentos de análisis* [*How to assess the quality of education based on ICT: guidelines and analysis tools*]. Barcelona: Editorial GRAÓ.

Crosby, P. B. (1979). *Quality is free: The art of making quality certain.* New York, USA: Nal Penguin Inc.

Dick, W., Carey, L., & Carey, J. O. (2005). *The systematic design of instruction*. Boston, MA: Allyn & Bacon.

Elissavet, G., & Economides, A. A. (2003). An evaluation instrument for hypermedia courseware. *Education Technology & Society*, *6*(2), 31-44.

ENQA (European Network for Quality Assurance in Higher Education). (2005). *Standards and guidelines for quality assurance in the European higher education area*. Retrieved October 19, 2014, from http://www.enqa.net/files/ENQA%20Bergen%20Report.pdf

Gagné, R. M., Briggs, L. J., & Wager, W. W. (1992). *Principles of instructional design.* Orlando, FL: Harcourt Brace Jovanovich College Publishers.

Garvin, D. A. (1992). *Managing quality: The strategic and competitive vision*. Rio de Janeiro, Brazil: Qualitymark Ed.

Godbole, N. S. (2005). *Software quality assurance: Principles and practice*. Oxford, UK: Alpha Science International Ltd.

Guri-Rosenblit, S. (2005). 'Distance education' and 'e-learning': Not the same thing. *Higher Education*, *49*(4), 467-493.

Hadjerrouit, S. (2007). Applying a system development approach to translate educational requirements into e-learning*. Interdisciplinary Journal of Knowledge and Learning Objects, 3,* 107-134*.* Retrieved November 5, 2011, from http://www.ijello.org/Volume3/IJKLOv3p107-134Hadj296.pdf

Humphrey, W. S. (1989). *Managing the software process*. Pittsburgh, PA: Addison-Wesley.

IHEP (The Institute for Higher Education Policy). (2000). *Quality on the line: Benchmarks for success in internet-bases distance education*.

ISO (International Standardization Organization). (2009). *ISO/IEC 19796-3:2009. Information technology – learning, education and training – Quality management, assurance and metrics – Part 3: Reference methods and metrics*. International Standardization Organization.

Knewton. (2012). The state of digital education infographic. *Learning Solution Magazine*. Retrieved October 21, 2014, from http://www.knewton.com/digital-education/

Khan, H. B. (2004). The people-process-product continuum in e-learning: The e-learning P3 model. *Issue of Educational Technology*, *44*(5), 33-40.

Litto, F. M., & Formiga, M. (2009). *Educação a distância: O estado da arte* [*Distance Learning: State of art*]. São Paulo, SP: Pearson Prentice Hall.

Martínez, D. R., García, F. B., González, E. E., Molina, P. G., Jorge, A. H., Nuñez, J. L. M., et al. (2011). *Gestión de proyetos de E-learning* [*Project Management of E-learning*]. México: Alfaomega Grupo Editor.

Marshall, S., & Mitchell, G. (2002).  An e-learning maturity model?  *Proceedings of ASCILITE Australasian Society for Computers in Learning in Tertiary Education*.  Retrieved May 03, 2013, from http://ascilite.org.au/conferences/auckland02/proceedings/papers/173.pdf

MarylandOnline. (2011). *Quality matters rubric standards 2011-2013*. Retrieved October 19, 2014, from http://www.qmprogram.org.

MEC/INEP (Ministério da Educação e Cultura / Instituto Nacional de Estudos e Pesquisas Educacionais Anisio Teixeira) [Ministry of Education and Culture / National Institute for Educational Studies AnisioTeixeira]. (2009). *Censo da Educação Superior 2008* [*Census of Higher Education 2008*].

MEC/SEED (Ministério da Educação e Cultura/Secretaria de Educação a Distância) [Ministry of Education and Culture / Department of Distance Education]. (2007). Referenciais da Qualidade para a Educação a Distância [*Quality benchmarks for distance education*]. Retrieved October 19, 2014, from http://portal.mec.gov.br/seed/arquivos/pdf/legislacao/refead1.pdf

Moore, M., & Kearsley, G. (2011). *Educação a distância: Uma visão integrada* [*Distance education: An integrated view*]. São Paulo, SP: Cengage Learning.

Nunes, V. B., Albernaz, J. M., & Nobre, I. A. M. (2009). Avaliação de cursos a distância [Evaluation of distance education courses]. *Anais do VI Congresso Brasileiro de Ensino Superior a Distância* [*Proceedings of the VI Brazilian Congress of Distance Education*], *1*(1), 1–10.

Pawlowski, J. M. (2007). The quality adaptation model: Adaptation and adoption of the quality standard ISO/IEC 19796-1 for learning, education, and training. *Educational Technology & Society*, *10*(2), 3-16.

Pera, S. M., Cervera, M. G., & Barado, S. I. (2007). E-Tutoria: Uso de las tecnologias de información y comunicación para la tutoria acadêmica universitária [E-Tutoring: Use of information and communication technologies for academic university tutoring]*, Revista Electrónica Teoría de la Educación* [*Electronic Magazine of Educational Theory*]*, 8(2)*, 31–54.

Perkins, J. E. P. (2008)*. Una introducción a la educación a distancia* [Introduction to distance education]. Buenos Aires, BA: Fundo de Cultura Económica.

Piva Júnior, D., Pupo, J. R. S., Gamez, L., & Oliveira, S.H.G. (2011). *EAD na prática: Planejamento, métodos e ambiente de educação online* [*EAD in practice: Planning, methods and online education environment*]. Rio de Janeiro, RJ: Elsevier.

PMI (Project Management Institute). (2008). *A guide to the project management body of knowledge (PMBOK®)*. PA: Project Management Institute.

Pressman, R. S. (2011). *Engenharia de Software: Uma Abordagem Profissiona.* [Software Engineering: A Professional Approach]. Porto Alegre, RS: AMGH Editora Ltda.

Rapchan, F. J. C., Cury, D., Menezes, D., & Falbo, R.A. (2002). Um Modelo de Qualidade de Processo para Cursos a Distância Mediados pela Internet [A model of quality for distance learning courses mediated by Internet]. *Proceedings of Second Simpósio Brasileiro de Qualidade de Software (SBQS), 1,* 1-15.

Rekkedal, T. (2006). State of the art report on distance learning and e-learning quality for SMEs. *EU Leonardo project, E-learning Quality for SMEs: Guidance and Counselling, May 2006,* 1-27. Retrieved November 11, 2011, from http://nettskolen.nki.no/in_english/elq-sme/ELQ-SMEStateofArt.pdf

Rossi, R. (2013). *eQETIC: Modelo de qualidade para soluções educacionais digitais* [eQETIC: Quality Model for Digital Educational Solutions]. São Paulo, SP: Editora Mackenzie.

Rossi, R., & Mustaro, P. N. (2012). Applying quality approaches in ICT-based educational products. *Proceedings of Informing Science & IT Education Conference (InSITE) 2012,* 249-264. Retrieved March 05, 2015, from http://proceedings.informingscience.org/InSITE2012/InSITE12p249-264Rossi0078.pdf

Shelton, K. (2011). A review of paradigms for evaluating the quality of online education programs. *Online Journal of Distance Learning Administration*, *4*(1), 1-9.

Sommerville, I. (2003). *Engenharia de Software* [Software Engineering]. São Paulo, SP: Pearson Education do Brasil.

West, C. K., Farmer, J. A., & Wolff, P. M. (1991). *Instructional design: Implications from Cognitive Science*. Boston, MA: Pearson Custom Publishing.

# Biographies

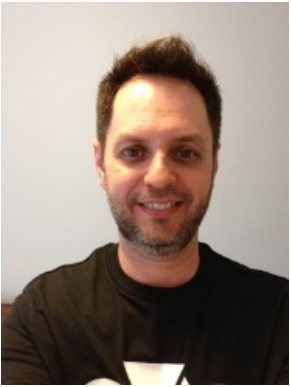

**Rogério Rossi** received his B. S. in Mathematics from the University **Center** Foundation Santo André; he also has a M.S. and Ph.D. in Electrical Engineering, both by Mackenzie Presbyterian University. He is in a Postdoctoral Program at the University of São Paulo developing research that is related to Complex Systems, Big Data, and the Internet of Things (IoT).

He is an Adjunct Professor for Information Technology and Computer Science courses of graduate and undergraduate programs in São Paulo. He has done research on the fields of software quality and quality for digital educational solutions, and he also has some publications on this area.

He is a member of IACSIT (International Association of Computer Science and Information Technology), and he worked as a reviewer for I$^n$Site Conferences 2013 and 2015, and e-Skills Conference 2014; he also presented his papers in the I$^n$Site Conferences in Montreal, Canada (2012) and Porto, Portugal (2013).

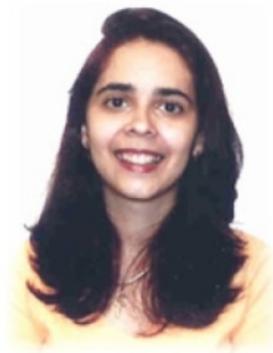

**Pollyana Notargiacomo Mustaro** was graduated in Pedagogy by the University of São Paulo, an institution where she also earned the title of Master and Doctor of Education. She is currently Professor at Mackenzie Presbyterian University, where she develops activities for Research and Teaching at the Computer Science College and Electrical Engineering Postgraduation Course. Among her areas of research, the following themes stand out: Instructional Design, Learning Objects Theory, Learning Styles, Distance Learning, Podcasts, Social Media Approaches and Technological Tools, Social Network Analysis, Hypertext Theory, Serious Games, Game Culture Studies, and Narratology.